# The Return on Investment in AI Ethics: A Holistic Framework

Marialena Bevilacqua
University of Notre Dame
mbevilac@nd.edu

Nicholas Berente
University of Notre Dame
nberente@nd.edu

Heather Domin
IBM
hesill@us.ibm.com

Brian Goehring
IBM
goehring@us.ibm.com

Francesca Rossi
IBM
Francesca.Rossi2@ibm.com

## Abstract

*We propose a Holistic Return on Ethics (HROE) framework for understanding the return on organizational investments in artificial intelligence (AI) ethics efforts. This framework is useful for organizations that wish to quantify the return for their investment decisions. The framework identifies the direct economic returns of such investments, the indirect paths to return through intangibles associated with organizational reputation, and real options associated with capabilities. The holistic framework ultimately provides organizations with the competency to employ and justify AI ethics investments.*

## 1. Introduction

The rapid innovation of ever-more powerful artificial intelligence (AI) technologies is driving benefits for organizations in terms of productivity, efficiency, and opportunity. Organizations that seek to adopt AI for these benefits need to be aware of possible ethical issues, such as those associated with bias, fairness, privacy, misinformation, fraud, and labor (Fjeld et al., 2020; Floridi et al., 2018; Kieslich et al., 2022). To address these issues, it is critical that organizations invest in AI ethics programs to deal with the unintended consequences of AI technologies, but it may be unclear what they receive in return. One study found that participants from organizations across sectors struggled with justifying the "additional time and resource costs associated with 'pro-ethical design', especially when there is no clear return on investment (Morley et al., 2021, p. 413)." While there is an abundance of research in the academic, technical, and corporate literature on AI ethics in general and in specific areas (e.g., bias, explainability, etc.), there is currently no method or tool to justify such AI ethics investments.

Motivated by this need, we propose a framework for establishing the return on investment (ROI) for organizational investments in AI ethics. We summarize the sorts of investments that organizations could make in AI ethics, characterize three types of ROI (economic, intangible, real options), and relate them in an original framework that we then apply to example AI ethics scenarios. We conclude with a discussion of AI ethics, as well as the need for its continual investment and methods for measuring ROI.

## 2. AI Ethics

Although AI has had a rebirth in recent years, it is not a new technology. Rather, AI has been around for a long time and has supported the frontier of computing for more than half a century – wave upon wave of innovations have been continually generated and applied in various situations (Berente et al., 2021). The current incarnation of AI exploits a variety of techniques, especially those based on machine learning approaches, that are used broadly across industries (Haenlein & Kaplan, 2019).

The pervasive use of AI technologies brings many benefits to people, society, and industry sectors (Davenport & Ronanki, 2018). However, the scale and pace at which AI technologies are deployed also raise concerns among a variety of stakeholders regarding potential ethics issues, particularly in high-stakes decision environments (Coeckelbergh, 2020). Examples of AI ethics issues are related to fairness, explainability, transparency, robustness, privacy, accountability, misinformation, value alignment, labor replacement, harmful content generation, and deep fakes, and they are especially concerning in autonomous AI systems (Fjeld et al., 2020; Floridi et al., 2018; Kieslich et al., 2022). In the past decade, AI stakeholders have worked to identify and mitigate these issues, with several complementary mechanisms such as principles, best practices, guidelines, software tools, playbooks, educational efforts, governance, standards, audits, certifications, and regulations (Fjeld et al., 2020; Floridi et al., 2018; Kieslich et al., 2022; Jobin et al., 2019). The most active organizations are those that build AI technologies, policy makers, standards bodies, AI research associations, and multi-stakeholder organizations.

For example, the OECD published AI principles to promote trustworthy AI, that have been endorsed by multiple countries (OECD, 2023). The European Commission published the guidelines for trustworthy AI in Europe (European Commission, 2023) and the European Parliament is currently discussing the AI Act, an AI regulation proposal (European Parliament, 2023). Organizations such as the Partnership on AI (2023), the Global Partnership on AI (GPAI, 2023), the World Economic Forum, and the United Nations have all published principles and guidelines to ensure a responsible approach to the adoption of AI. Individual AI organizations have built AI ethics frameworks. As an example, IBM has published a set of AI ethics principles and trustworthy AI pillars, an internal risk assessment process for every client offering, and playbooks to help AI developers embed ethics by design in AI systems. In addition, IBM has also created software tools to identify and mitigate risks, AI ethics educational material for all employers, and an AI ethics board to ensure coordination and governance for all these activities (IBM, 2023; World Economic Forum, 2021). For example, the IBM software tool AI Decision Coordination (Baudel et al., 2023) provides an interactive experience to define custom metrics and optimally allocate decision tasks between humans and AI. These capabilities may help engender consensus building among the stakeholders involved in the decision-making process and can help assess potential impacts and the optimal level of human control and oversight.

While publishing or endorsing AI ethics principles may be expedient, translating those principles into concrete actions requires significant investments. This is particularly challenging because AI technology evolves very rapidly, so AI ethics frameworks need to stay abreast of new issues as they arise from the "ever-evolving frontier of computational advancements (Berente et al., 2021, p. 1433)." Such investments need to be justified to be fully supported within an organization.

The typical way that organizations justify technology investments, including AI, is by calculating their "return on investment" (ROI) (Bartel, 2000; Phillips, 1994; Hoffman & Fodor, 2010). However, computing the return on AI ethics investments is challenging because of the presence of multiple types of returns, often intangible, that concur in both direct and indirect ways to the overall tangible return. This paper puts forward a holistic framework to identify the return on AI ethics investments.

## 3. Return on Investment (ROI)

Organizations commonly justify investments by calculating ROI using the standard mathematical definition focusing on economic return (Bartel, 2000; Phillips, 1994; Hoffman & Fodor, 2010). While this is adequate for some situations, investments in AI ethics require an expanded view of ROI and also take into consideration the intangible and indirect returns. Therefore, we present three approaches to ROI: traditional (economic), intangible (reputational), and real options (capabilities).

Stakeholders are defined as anyone who "can affect or is affected by the achievement of the organization's objectives" – and different stakeholders think of return through different units of analysis (Freeman et al., 2010). Therefore it is important to consider both the unit of analysis and the relevant stakeholders in calculating the ROI.

### 3.1. Traditional ROI: Economic Impact

Traditionally, ROI is calculated by dividing a measure of economic return by the cost of an investment. What constitutes the numerator and denominator depends on whether ROI is intended to measure return on the organization overall, or return on specific investments within an organization.

Computing ROI at an organizational level considers the organization as the unit of analysis and managers and shareholders as the primary stakeholders. Organizational-level ROI, from an accounting standpoint, is calculated as the ratio of net operating profit to the net book value of assets (Richard et al., 2009). From a standpoint of investment finance stock performance, stock return is often deemed a superior measure of return (Mitchell et al., 1997; Jacobson, 1987) and replacement costs of assets is thought to be superior to book value (i.e. Tobin's Q; Landsman & Shapiro, 1995).

Within an organization, ROI is calculated to assess particular expenditures, such as new subunits, projects, capital equipment, etc. For subunits, for example, organizations often take the net present value over time of a unit's cash contribution as the numerator (discounted cash flows over operating assets; see Richard et al., 2009). Subunits can also parse the denominator to look at different expenditures such as research and development (R&D) and capital equipment (Hsieh et al., 2003). ROI is used to assess project performance as a way of assessing and comparing contributions of all sorts of projects such as software projects, marketing campaigns, information technology projects, and employee healthcare programs (Kwak & Ibbs, 2000; Bockle et al., 2004; Mehra et al., 2014; Kumar & Mirchandani, 2012; Menachemi et al., 2006).

## 3.2. Intangible ROI: Reputational Impact

Oftentimes investments do not have a clear and direct, traceable impact on financial outcomes, yet still provide value to the organization and its constituents. Often this value is intangible and impacts social, cultural, or psychological aspects of key stakeholders. Intangible outcomes are not easily quantifiable but are nevertheless critical to organization citizenship, competitiveness, and survival (Ballow et al., 2004).

There are numerous forms of intangible ROI. Perhaps the most well-established management perspective for intangibles is "corporate social responsibility" (CSR), in which organizations are committed to generating value for a variety of stakeholders such as customers, employees, the local community, and society (Watts & Holme, 1999). Areas of CSR include, but are not limited to, employee relations, human rights, corporate ethics, and the environment (Moir, 2001). Multiple ratings were created to compare an organizations' CSR efforts and achievement (Márquez & Fombrun, 2005). Environmental, social, and governance (ESG) is a recent movement in the CSR tradition that encompasses an organization's ability to incorporate societal concerns in their business operations (Gillan et al., 2021; Corporate Finance Institute, 2023). Strong CSR and ESG ratings may improve sales and reputation and can decrease the likelihood of customer churn and employee turnover, but these downstream benefits are difficult to trace. However, there is some evidence that programs such as CSR and ESG can positively influence financial performance and business value (Friede et al., 2015; Wong et al., 2021). In order to understand, manage, and report the economic value in conjunction with social impacts, a technique called Social Return on Investment (SROI) was developed (Millar & Hall, 2013). This technique is used globally to measure the social and economic value of social enterprises (Nicholls, 2007; SROI Network, 2011; Ravulo et al., 2019).

Organizations also assess the impact of investments on their culture, which is particularly salient for employee stakeholders. Organizational culture refers to the shared beliefs and values guiding the thoughts and behaviors of its members, which can be assessed both qualitatively and quantitatively (Cooke & Rousseau, 1988). The relationship between corporate culture and performance has long been established, but widespread use of tools and methods to measure this impact are a more recent development (Shahzad et al., 2012). There are a variety of instruments developed to capture different aspects of an organization's culture (Denison & Neale, 2000; Roos & Van Eeden, 2008; Denison, 1984).

Many intangible outcomes focus on customer stakeholders including customer satisfaction and brand loyalty (Ahmed et al., 2014; Bloemer & Lemmink, 1992; Al-Msallam, 2015; Awan & Rehman, 2014). Measures of these constructs are either attitudinal or behavioral. Attitudinal measures utilize stated preferences, whereas behavioral measures utilize actual customer activity (Mellens et al., 1996). Attitudinal measures historically involved surveys (Guest, 1942; Traylor, 1981), but more recently can involve online assessments of sentiment. Behavioral measures include switching matrices (Mellens et al., 1996), word of mouth (WOM) (Kumar et al., 2007), and net promoter scores (Reichheld, 2003).

## 3.3. Real Options and ROI as Capabilities

Options theory is rooted in the work of Black and Scholes (1973) and Merton (1973) on financial options that involve capitalizing on upside potential while minimizing downside financial risk with partial investments. A financial option involves a small investment for the right to invest in the future. "Real options" are similar, but do not refer to financial instruments. Real options refer to small investments made by organizations that generate future flexibility – they position managers to make choices that capitalize on future opportunities (McGrath et al., 2004). Real options reasoning promotes proactive learning that builds the capabilities necessary for organizations to execute on novel projects in an uncertain future. Instead of only utilizing the knowledge present at the beginning of the project, organizations can effectively utilize knowledge acquired throughout the duration of the project to improve outcomes (McGrath et al., 2004). Additionally, organizations can incrementally approach major investments in disciplined stages, sequencing smaller investments to ensure that opportunities with significant upside are pursued despite initial uncertainty. This staged investment approach maintains divisibility among projects to avoid correlation and compounding risk.

Real options theory can be applied to scenarios where return on investment is uncertain. Real options reasoning is used to assess and justify investment in novel technology projects that involve building knowledge capabilities and skills, as well as technical infrastructures that can be used as a platform for future flexibility and innovation (Woodard et al., 2013; Sambamurthy et al., 2003). STAR© (strategic technology assessment review), exemplifies a real options approach to evaluate limited investments in technological assets to avoid further investments (McGrath & MacMillan, 2000). Return of real options is calculated as a function of the claim on the potential upside of a series of investments, less the cost of those investments. A defining characteristic of real options reasoning is that those in decision-making positions can maintain fiduciary responsibility even when making sometimes uncertain, aggressive investments.

## 4. A Holistic Framework

AI ethics principles and guidelines are indeed important and a good start, but organizations need to do more to ensure that they stay abreast of new issues as they arise from the ever-evolving frontier of AI (Berente et al., 2021, p. 1433). They must make investments that ensure the ethical use of AI technologies. Such investments could involve employee education and training, building compliant software tools, defining risk assessment and governance frameworks, or creating a center of excellence (COE), all of which establish and enforce ethical AI practices. Investments in AI ethics are clearly important, but can be quite costly. Therefore, we propose that organizations interested in determining the impact of their investments in AI ethics apply measures of ROI.

As AI adoption becomes ubiquitous, so has the desire to understand its impact (Frank et al., 2019; Young et al., 2019; Chatterjee et al., 2021). AI investments themselves can contribute significant ROI to organizations (Ashoori et al., 2023), but this is not typically distinguished from specific investments that assure ethicality in these efforts. Understanding the impact of specific investments in AI ethics is critical. Return, particularly around ethical issues, is always with respect to some stakeholder. As such, understanding the impacts of an organization is fundamentally a question about how their activities impact stakeholders (Freeman et al., 2010). The ethical approach of an organization – its culture of ethics – is inextricably linked to its view of its relationship with its stakeholders (Jones et al., 2007). To make sense of the impact of investments in AI ethics, it is therefore important to understand return with respect to particular stakeholders. There are three paths to understanding the impact of investments in AI ethics with regards to stakeholders: the direct path through economic return, and indirect paths through capabilities and reputation (see Figure 1). The Holistic Return on Ethics (HROE) framework presented in Figure 1 encompasses and describes the relationships, stakeholders, and possible returns that exist when organizations make investments in AI ethics. Next, we discuss each in turn.

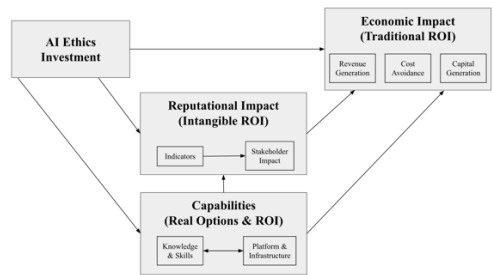

**Figure 1. HROE: A Framework for AI**

The traditional financial ROI measures the direct relationship between an investment in AI ethics and its consequent economic return in terms of cost savings, revenue generation, or reduction of cost of capital. Intangible ROI captures proximal reputational impacts of investments in AI ethics such as stakeholder trust, which are likely to eventually lead to proximal economic ROI. Finally, we consider the optionality of AI ethics investments in terms of building proximal AI ethics related capabilities, which provide the options to the organization for subsequent flexibility and cost savings. To operationalize our framework, we provide formal models with hypothetical illustrations. While these illustrations focus on AI, the framework is also useful for determining ethics investments required for other types of technologies.

### 4.1. The Economic Impact of AI Ethics Investments

As previously mentioned, the traditional financial ROI measures the direct relationship between an investment in AI ethics and its economic return from relevant stakeholders. This return may come in the form of revenue generation, cost savings, or reduction of cost of capital, or even a combination of the three. In Table 1, examples of the three types of returns one may experience due to an investment in AI ethics are provided. The table depicts the additional revenue and capital an organization may generate and the costs they will avoid in respect to the stakeholder referenced. While the type and magnitude of the return, as well as the stakeholder providing the return may vary, the organizational intention of investing in AI ethics for positive economic impacts is invariable.

Organizations may be inclined to make investments in AI ethics in part due to the potential revenue that can be generated from each stakeholder. For example, suppose an organization invests in debiasing their AI systems that are trained on outdated and biased data. As a result, previously untapped customers may be more inclined to purchase the organizations' goods or services, enlarging its customer base and furthermore increasing its revenue.

Organizations can also experience beneficial economic impact in the form of cost avoidance by making investments in AI ethics. This may result in the organization's avoidance of governmental fines and may lower the cost of governmental compliance. In 2021, Facebook ceased the operations of its Face Recognition system due to privacy and transparency concerns, which it began operating prior to clear regulatory guidance concerning facial recognitional technology in society (Pesenti, 2021). In doing so, Facebook avoided costs associated with additional perception management and positively impacted their likelihood of revenue generation.

Lastly, organizations may receive return on AI ethics investments via capital generated by various stakeholders. For example, when organizations make investments in novel AI ethics inventions, the probability of receiving intellectual property (IP) and its associated rights from the government will increase. The significance and relevance of this return on AI ethics investments is exemplified in Stanford's 2023 AI Index report that suggests that IP is currently a topic heavily discussed and debated in both academia and industry (Maslej & Lynch, 2023). Not only could obtaining IP rights directly impact the organization's revenue due to their sole ownership of the invention, but it could also contribute to the organization's capital by encouraging the investor community to continue purchasing the organizations' shares.

**Table 1. Economic Returns**

|  | Revenue Generation | Cost Avoidance | Capital Generation |
|---|---|---|---|
| Shareholder |  | -Cost of investor relations | -Share purchases |
| Government | -Research grants<br>-Funding | -Cost of compliance<br>-Fine avoidance<br>-Legal fees | -Intellectual property & patents<br>-Business loans |
| Employee |  | -Cost to recruit & retain<br>-Cost to onboard new employees<br>-Overhead costs<br>-Resource costs for appropriately automated processes |  |
| Customer | -Consulting services<br>-Partner sales<br>-Revenue protection (core products & services)<br>-More employee purchases<br>-Expansion of customer base | -Customer acquisition and retainment costs<br>-Litigation costs<br>-Service, maintenance, recall, etc. costs |  |

The following equation determines the present value of an AI ethics investment's economic return across N years. N denotes the current year as 1, and increases as years progress. The numerator represents the net economic return at time j ($R_j^e$), which is a function of the initial investment ($I_t$) at time (t). This term is discounted by subtracting the firm's cost of capital from 1, and then multiplying this value by a power of j-t. Mathematically, this is represented as $(1-\alpha)^{j-t}$ and is equivalent to the discounted present value $(\alpha_{j-t})$, at time (t).

$$ROI_{t,N} = \frac{(\sum_{j=t}^{t+N} \alpha_{j-t} R_j^e(I_t)) - I_t}{I_t}$$

By way of an illustrative (toy) example, imagine that a mid-sized organization is evaluating whether to implement a platform to manage risks associated with AI technologies, in hopes that they will avoid unnecessary fines in the future. Let's assume that this company has revenues of $100 million and the investment in the platform costs $1 million dollars. Under the proposed European Artificial Intelligence Act, a fine for non-compliance with this Act could be as much as 6% of an organization's revenue for a total potential fine of $6 million (European Commission, 2021). Let's assume the Act is enforced in year two, the year of the hypothetical fine avoidance, and the firm's cost of capital is 10% (which informs its discount rate from year 2 (j) to year 1 (t), $(\alpha_{j-t})$). Therefore, the economic return of this investment will be calculated by multiplying 90% (1 – 0.10) by 6% of $100 million ($5.4 million). Subtracting the initial $1 million investment ($4.4 million), and dividing by that $1 million, the ROI on the AI risk management platform is 4.4, or 440%. We use this oversimplified example in the following sections to illustrate how to quantify reputational and capabilities-related returns.

### 4.2. The Intangible Impact of AI Ethics Investments

While the economic return on an investment is crucial in understanding its monetary value, the intangible return on an investment is crucial in understanding its holistic value. In considering the intangible ROI, an organization will not only recognize previously unknown value that exists among its stakeholders, but may also encounter improvements in economic ROI as a result. As demonstrated in Figure 1 and exemplified in Table 2, improvements in indicators generate intangible impacts, in which relevant stakeholders formulate perspectives and reputations of an organization that have the potential to generate additional economic return. Eventually, these intangible and reputational impacts may lead to an increase in economic return in the form of revenue generation. Organizations can interpret the intangible impact of AI ethics investments through various indicators, as well as the economic return attributed to the indicators.

Intangible returns can be referred to as reputational returns, driven by the socially responsible reputation that an organization's stakeholders form due to its investment in AI ethics. Furthermore, shareholders, the government, employees, and customers develop perspectives of organizations as a result of their indicators, leading to the ensuing stakeholder impact. As an example, IBM invests in technology ethics training for its ecosystem partners and calls for proposals through the Notre Dame-IBM Tech Ethics Lab to conduct "tangible, applied, and interdisciplinary research that addresses core ethical questions (2023; 2023)." Due to its investments in the social aspect of ESG such as community impact and equity, education, and social innovation, IBM's ESG score and its ability to retain employees may improve (Deloitte, 2023). In IBM's 2022 AI ethics report, it was found that "among employees, nearly 70% said they are more likely to accept a job offer from an organization they consider to be environmentally and socially responsible, and a similar dynamic impacts retention" (IBM 2022). Organizations that make direct investments in business and AI ethics contribute to the governance aspect of their ESG rating, further improving their overall score (Tang, 2019).

In recent years, the field of natural language processing (NLP) has given rise to large language models (LLMs), such as OpenAI's ChatGPT, that can generate human-like output. Although many are intrigued by the societal benefits of LLMs, these generative AI models pose environmental harms such as a worsening carbon footprint and an increase in pollution (Rillig et al., 2023). By investing in the reduction of carbon emissions attributed to AI technologies, organizations can improve their ESG ratings. If organizations make investments to improve the environmental, social, or governance components of their carbon footprint, they will not only impact the ESG score, but also potentially impact the perspectives of a variety of stakeholders. More generally, through improvements in their indicators, organizations will impact stakeholder perceptions, thus increasing brand loyalty, trust, and satisfaction.

Organizations may also be interested in audits that substantiate the ethicality of their AI systems. O'Neil Risk Consulting & Algorithmic Auditing (ORCAA), for example, is a consultancy that helps organizations manage and audit algorithmic risks related to fairness, bias, and discrimination (O'Neill, 2023). By requesting and conducting successful audits with consultants such as those at ORCAA, organizations may experience an increase in their recruitment metrics, as Millennials and Gen-Z are motivated to work for "socially responsible employers" that "prioritize purpose (Aziz, 2021)."

The ethical development and use of AI systems may directly impact an organization's economic return. In addition, organizations may also unintentionally receive indirect return through media coverage of their artifacts. IBM developed a variety of AI ethics tools such as AI Fairness 360, Adversarial Robustness 360, and AI Explainability 360. In doing so, the organization received media coverage from technology and business sources such as VentureBeat and TechTarget, promoting the tools' mitigation of advertising bias and the organizations' donations to the Linux Foundation (Lawton, 2022; Labbe, 2020). The topics covered by the media are likely to lead to positive affect among shareholders, improved morale among employees, and increased trust in the organization's commitment to ethics.

Table 2. Reputational Returns

|  | Indicators | Stakeholder Impact |
|---|---|---|
| Shareholder | -ESG & CSR<br>-Successful audits<br>-Certifications<br>-Media coverage<br>-Investment<br>-Third party endorsements/ mentions<br>-Labor/citizen actions<br>-Direct leads<br>-Partnerships<br>-Memberships<br>-Legal actions<br>-Regulatory actions<br>-Employer ratings | -Propensity to invest<br>-Positive affect |
| Government |  | -Political position & power |
| Employee |  | -Morale<br>-Retention/attrition<br>-Recruitment |
| Customer |  | -Brand loyalty<br>-Trust<br>-Satisfaction |

We can extend our model of direct return to also include indirect, intangible, reputational return ($R_j^r(I_t)$).

$$ROI_{t,N} = \frac{\left(\sum_{j=t}^{t+N} \alpha_{j-t}\left[R_j^e(I_t) + R_j^r(I_t)\right]\right) - I_t}{I_t}$$

Assuming the same fine avoidance from the example in the previous section ($6 million), assume positive media coverage in year two which the organization values at $0.5 million (discounted present value is $0.45 million). Thus the combination of economic ($5.4 million) plus the intangible ($0.45 million) less the investment ($1 million), results in a total return of $4.85 million, which represents an ROI of 485%.

### 4.3. The Return on AI Ethics Investment Options as Capabilities

Employing a real options approach to investments in AI ethics allows organizations to reason through the value of additional capabilities. The real options method proposes a staged-investment approach that allows organizations to continually learn, attain flexibility in making decisions, and ultimately save costs. Consequently, the relevant stakeholder's knowledge, skills, platform, and infrastructure capabilities are broadened. Through proactive learning, stakeholders gain knowledge and skills that they may not have otherwise acquired. To quantify the return on an investment made using the real options approach, one can use the formulation provided by McGrath and MacMillan (2000) by acquiring the cumulative returns, initial investment costs, and subsequent stepped investments in real options.

Table 3 details examples of the capabilities an organization may obtain after making AI ethics investments using a real options approach. When an organization invests in AI governance, this action implies the prioritization of embodying ethical standards set by itself, its stakeholders, and relevant regulation. In doing so, the organization will acquire knowledge and skills relative to each of their stakeholders in areas such as risk assessment, opportunity recognition, regulatory compliance, and policy. More specifically, the employees will gain a deeper understanding of the regulatory compliance that is applicable to each individual stakeholder. The acquired regulation and compliance knowledge may require the organization to strengthen its platform and infrastructure perhaps through the creation and implementation of regulatory technology (RegTech). Investments in AI ethics can also lead to further ensure the organization consistently meets regulatory standards (Gentile, 2022).

AI products are ubiquitous and organizations responsible for their creation must work to ensure their ethical development and utilization. An organization can achieve ethical use among its customers by investing in AI products that incorporate ethical insights throughout the products' development. As an example, employees managing the construction of smart home appliances such as Amazon's Alexa or Netflix's streaming service must guarantee that the customers' data is protected, the products' quality is high, and more importantly, that the customer is safe. In building a platform that embodies these criteria, the organization can apply these ethical product development practices to all of its current and future products. The capabilities an organization gains from an investment in ethical product development will further result in opportunities for possible product improvements, the mitigation of identified product issues, and possibly a greater awareness and thus improvement in the organization's employee and customer culture.

Table 3. Returns as Capabilities

|  | Knowledge & Skills | Platform & Infrastructure |
|---|---|---|
| Shareholder | -Risk assessment | |
| Government | -Opportunity recognition -Regulatory compliance -Influence policy -Mitigate identified issues -Culture | -Software -Regulatory technology |
| Employee | | -Risk assessment tooling -Product & data management -Regulatory technology |
| Customer | | -Improved product quality and safety |

To quantify the returns as capabilities of an AI ethics investment we apply the following model. The first two elements of the numerator in this equation represent the economic ($R_j^e(I_t)$) and reputational returns ($R_j^r(I_t)$) from the previous models. The final element represents the returns as capabilities given a total number of options (m). This new sum also accounts for the net return ($R_{t,k}^c$) from real options in the numerator and the cost of the investment ($I_t^c$) for the options investment in the denominator. The $\gamma$ variable in the last term of the numerator represents the fraction of the allotted funds invested for a particular investment ($I_t^c$). This term can take values between 0 and 1, since employing the real options technique means that the cost of an initial investment may only be part of the total funds allocated for the investment. Additionally, the sum of all $\gamma_k$ must equal 1, as the total will be comprised of each individual options investment. Following is the HROE equation:

$$ROI_{t,N,m} =$$

$$\frac{\left(\sum_{j=t}^{t+N} \alpha_{j-t}[R_j^e(I_t) + R_j^r(I_t) + \sum_{k=1}^{m} R_{t,k}^c(\gamma_k I_t^c)]\right) - (I_t + I_t^c)}{I_t + I_t^c}$$

Applying the assumptions from the previous examples (investment of $1 million, return of $4.85 million), assume that in the first year the organization's managers realized that they could use the risk management platform to also save about $0.5 million ($R_{t,k}^c$) in the second year in software testing, but that this additional capability would cost an additional $0.3 million ($I_t^c$) investment in the first year. Although investments increase $0.3 million, returns (discounted) total to $0.45 million. The $0.3 million investment subtracted from the $0.45 million return is $0.15 million. Thus a total of $5 million ($4.85 + $0.15 million) divided by $1.3 million ($1 + $0.3 million) is a total $ROI_{t,N,m}$ of 3.85, or 385%.

## 5. Discussion

By now, many organizations using or building AI technology – or incorporating it into their business models, products, or operations – are realizing that they need to invest in AI ethics. The kind and scale of investment depends on the organization's type, size, goals, and business units, as well as its role as an AI provider or consumer (a distinction that is increasingly blurring into more of a spectrum). Different investments have different implications and may result in one of the three previously mentioned ROIs more than the others. For example, if an organization invests in data stewardship to protect the privacy and information of the users of its AI technologies, it will experience economic return in the form of revenue. On the other hand, if an organization invests in the mitigation of bias and thus promotion of fairness in using their AI technologies, it will most likely receive intangible return. Finally, organizations will mostly receive return in the form of optionality when making investments in areas such as the enhancement of technological software or human resource operations. This framework is novel and unique, yet is limited in that it is not backed by data. In future research, we hope to collect and utilize data to exemplify the practicality and applicability of this framework.

The framework provided in this paper is tailored to scenarios in which organizations aim to determine their overall holistic return on AI ethics investments. However, this framework is generalizable to other contexts involving various types of investments due to its holistic nature. By acknowledging the existence of, and providing measures for the economic, reputational, and capabilities-related impacts of an investment, the true total return on investments in AI ethics can now be identified, addressed, referenced, and further developed in future research. In particular, as data from society at large – consumer, citizens, and others – increasingly contribute to the value of AI investments themselves, organizations may need to broaden the set of stakeholders considered in assessing the value of the returns on these investments, which this framework may help to determine. Most importantly, organizations now possess greater clarity of the benefits they can incur, relative to each stakeholder, by investing in AI ethics.